\input epsf
\input harvmac
\noblackbox
\newcount\figno
\figno=0
\def\fig#1#2#3{
\par\begingroup\parindent=0pt\leftskip=1cm\rightskip=1cm\parindent=0pt
\baselineskip=11pt
\global\advance\figno by 1
\midinsert
\epsfxsize=#3
\centerline{\epsfbox{#2}}
\vskip 12pt
\centerline{{\bf Figure \the\figno:} #1}\par
\endinsert\endgroup\par}
\def\figlabel#1{\xdef#1{\the\figno}}

\font\cmss=cmss10
\font\cmsss=cmss10 at 7pt
\def\rlx{\relax\leavevmode}
\def\inbar{\vrule height1.5ex width.4pt depth0pt}
\def\IC{\relax\,\hbox{$\inbar\kern-.3em{\rm C}$}}
\def\IN{\relax{\rm I\kern-.18em N}}
\def\IP{\relax{\rm I\kern-.18em P}}
\def\ZZ{\rlx\leavevmode\ifmmode\mathchoice{\hbox{\cmss Z\kern-.4em Z}}
 {\hbox{\cmss Z\kern-.4em Z}}{\lower.9pt\hbox{\cmsss Z\kern-.36em Z}}
 {\lower1.2pt\hbox{\cmsss Z\kern-.36em Z}}\else{\cmss Z\kern-.4em Z}\fi}
\def\narrowplus{\kern -.04truein + \kern -.03truein}
\def\narrowminus{- \kern -.04truein}
\def\narrowminussub{\kern -.02truein - \kern -.01truein}

\def\sqr#1#2{{\vcenter{\vbox{\hrule height.#2pt
            \hbox{\vrule width.#2pt height#1pt \kern#1pt
                  \vrule width.#2pt}\hrule height.#2pt}}}}

\def\wt{\widetilde}

\nref\rORIENT{
A. Sagnotti, in {\it Cargese '87, Non-perturbative Quantum
Field Theory}, ed. G. Mack et. al. (Pergamon Press, 1988) 521\semi G. Pradisi
and A. Sagnotti, Phys. Lett. {\bf B216} (1989) 59\semi
P. Horava, Nucl. Phys. {\bf B327} (1989) 461\semi
J. Dai, R. G. Leigh, and J. Polchinski, Mod. Phys. Lett. {\bf A4} (1989) 2073.}

\nref\rBPS{M. Bianchi, G. Pradisi, and A. Sagnotti, Nucl. Phys. {\bf B376}
(1992) 365\semi C. Angelantonj, M. Bianchi, G. Pradisi, A. Sagnotti, and Y.
Stanev, Phys. Lett. {\bf B385} (1996) 96 [hep-th/9606169]; Phys. Lett. {\bf
B387} (1996) 743 [hep-th/9607229].}

\nref\rPOL{E. G. Gimon and J. Polchinski, Phys. Rev. {\bf D54} (1996) 1667
[hep-th/9601038].}

\nref\rGJ{E. G. Gimon and C. Johnson, Nucl. Phys. {\bf B477} (1996) 715
[hep-th/9604129];
Nucl. Phys. {\bf B479} (1996) 285 [hep-th/9606176].}
\nref\rDP{A. Dabholkar and J. Park, hep-th/9607041;
Nucl. Phys. {\bf B477} (1996) 701
[hep-th/9604178];
Nucl. Phys. {\bf B472} (1996) 207 [hep-th/9602030]\semi
J. Park, hep-th/9611119.} 
\nref\rBS{M. Bianchi and A. Sagnotti, Phys. Lett. {\bf B247} (1990) 517;
Nucl. Phys. {\bf B361} (1991) 519.}
\nref\rBZ{J. D. Blum and A. Zaffaroni, Phys. Lett. {\bf B387} (1996) 71
[hep-th/9607019]\semi
J. D. Blum, Nucl. Phys. {\bf B486} (1997) 34 [hep-th/9608053].}
\nref\rGM{R. Gopakumar and S. Mukhi, Nucl. Phys. {\bf B479} (1996) 260
[hep-th/9607057].}

\lref\rFGEN{A. Sen, hep-th/9702165.}
\lref\rASH{A. Sen, Nucl. Phys. {\bf B475} (1996) 562 [hep-th/9605150].}
\lref\rVAFA{C. Vafa, Nucl. Phys. {\bf B469} (1996) 403 [hep-th/9602022].}
\lref\rEDCON{E. Witten, private communication.}
\lref\rSW{N. Seiberg and E. Witten, Nucl. Phys. {\bf B471} (1996) 121
[hep-th/9603003].}

\nref\rWITSIX{M. Berkooz, R. Leigh, J. Polchinski, J. Schwarz, N. Seiberg and
E. Witten, Nucl. Phys. {\bf B475}, (1996) 115 [hep-th/9605184].}
\lref\rASPINEW{P. Aspinwall, hep-th/9612108.}
\lref\rASPINK{P. Aspinwall, hep-th/9611137 and references therein.}
\lref\rNIKULIN{V. Nikulin, in {\it Proceedings of the International Congress of
Mathematicians}, Berkeley, 1986, 654.}
\lref\rASPMOR{P. Aspinwall and D. Morrison, in {\it Mirror Symmetry II}, ed. B.
Greene and S.-T. Yau (International Press, 1996) 703 [hep-th/9404151].}
\lref\rNSW{K. Narain, H. Sarmadi and E. Witten, Nucl. Phys. {\bf B279} (1987)
369.}
\lref\rFTHEORY{D. Morrison and C. Vafa, Nucl. Phys. {\bf B473} (1996) 74
[hep-th/9602114];
Nucl. Phys. {\bf B476} (1996) 437 [hep-th/9603161].}

\Title{\vbox{\hbox{hep--th/9703157}\hbox{MRI--PHY/P970304
IASSNS--HEP--97/23 NI 97017}}}{\vbox{\centerline{The Mirror
Transform of Type I Vacua in
Six Dimensions}}}
\centerline{ Ashoke Sen\footnote{$^{1}$}{E-mail: sen@mri.ernet.in,
sen@theory.tifr.res.in; on leave of absence from the
Tata Institute of Fundamental Research, Homi Bhabha Road,
Bombay 400005, INDIA.}}
\vskip 4pt
\centerline{\it Mehta Research Institute of Mathematics}
\centerline{\it and Mathematical Physics }
\centerline{\it  Chhatnag Road, Jhoosi, Allahabad 221506, INDIA}

\centerline{and}

\centerline{\it Isaac Newton Institute for Mathematical Sciences}
\centerline{\it University of Cambridge,  Cambridge CB3 0EH, U.K.}
\vskip 0.1in
\centerline{and}
\vskip 0.1in
\centerline{Savdeep Sethi\footnote{$^2$}{E-mail: sethi@sns.ias.edu} }
\medskip\centerline{\it School of Natural Sciences}
\centerline{\it Institute for Advanced Study}\centerline{\it
Princeton, NJ
08540, USA}

\vskip .3in

 We study certain compactifications of the type I string on $K3$. The three
topologically distinct choices of gauge bundle for the type I theory are
shown to be equivalent to type IIB orientifolds with different choices of
background anti-symmetric tensor field flux. Using a
mirror transformation, we relate these models to orientifolds with fixed
seven planes, and without any antisymmetric tensor field flux.
This map allows us to relate these type I vacua to
particular six-dimensional F theory and heterotic string compactifications.

\vskip 0.1in
\Date{3/97}


\newsec{Introduction}

String vacua in six dimensions with N=1 supersymmetry provide an interesting
arena for studying non-perturbative string dynamics. Many of the
difficulties that arise in four-dimensional compactifications, such as
space-time superpotentials, are absent. Among the simplest ways of constructing
six dimensional models are compactifications of the heterotic string, or type I
string on $K3$. Some of these theories have dual realizations in terms of
six-dimensional M or F theory compactifications.
Orientifolds provide another interesting class of string compactifications
in six dimensions.

An orientifold is
a generalized orbifold where the quotient group includes a usual space-time
symmetry together with world-sheet orientation reversal, and possibly
some other internal symmetry transformation \refs{\rORIENT,\rPOL}.
Orientifolds have proven useful for constructing models with a number of
interesting features such as extra tensor multiplets, or the analogue of
discrete torsion \refs{\rBPS-\rGM}. Type I string theory on $K3$ provides us
with a special class of orientifolds in six dimensions, since these models can
be
regarded as type IIB on $K3$ modded out by the world-sheet orientation
reversal symmetry, $\Omega$. By the usual duality between type I and
Spin(32)/$Z_2$  
heterotic string theory, these models are equivalent to
Spin(32)/$Z_2$ heterotic string theory on $K3$.

On the other hand,
a type IIB orientifold can often be related to a compactification of F theory
\rVAFA. An F
theory compactification is a particular background of the type IIB string
described in terms of an elliptically-fibered space, which we shall take to be
a Calabi-Yau manifold. The type IIB string is compactified on the base of the
Calabi-Yau manifold, with the variation of the complexified string coupling
specified by the variation of the complex structure modulus of the torus fiber.
Alternatively, we can view this background as type IIB on a space which has
positive first Chern class, but where seven-branes are used to construct a
consistent vacuum. In certain cases, there are regions of the
moduli space of the F theory compactification where the
dilaton-axion background
coincides with the background of a type IIB orientifold \refs{\rASH,\rFGEN}. In
these cases, we have two descriptions of the physics in terms of either F
theory, or the orientifold. In many of these cases, we expect F theory to
capture some of the non-perturbative physics of the orientifold. This is one of
the reasons F theory has proven so useful.
In cases where the F theory compactification has an $E_8\times E_8$
heterotic string dual, this relation provides a means of connecting the
orientifold to the $E_8\times E_8$ heterotic string.

The orientifolds which can be related to F-theory this way all have the
property that in any element of the quotient group, the world-sheet
parity transformation, $\Omega$, is always accompanied by $(-1)^{F_L}$,
the internal symmetry that changes the sign of all the left Ramond states.
This is due to the fact that in F-theory monodromies along closed curves
on the base must be elements of the SL(2,Z) self-duality group of type
IIB, and although neither $\Omega$ nor $(-1)^{F_L}$ is an element of
SL(2,Z), $(-1)^{F_L}\cdot\Omega$ is an element of SL(2,Z).
Thus, the type IIB orientifolds which correspond to compactification
of type I string theory cannot be directly related to F-theory, since
the former has $\Omega$ as one of the elements of the quotient group.
However, these two classes of orientifolds
can sometimes be related
by using T-dualities. For example, let us consider the
eight-dimensional orientifold constructed by quotienting type IIB on a
two-torus by $(-1)^{F_L}\cdot\Omega$, together
with an inversion of the torus. This particular
orientifold can be related to type I on the two-torus by T-dualizing both
circles of the torus \rASH.
Our discussion in this paper will focus on a generalization of this procedure
to six dimensions, {\it i.e.} studying
the relation between compactifications of the type I string on $K3$, and
orientifolds of the type IIB string on $K3$ which can be related to F-theory
vacua.

Understanding how to relate IIB orientifolds to type I vacua
in six and lower dimensions is of interest for several reasons.
Relating type I theories to the special class of
orientifolds which can be related to
F theory compactifications, allows us to see
relations between string vacua which are otherwise hard to determine. In
addition, even in six dimensions, there are regions of the moduli space of
heterotic strings on $K3$ for which neither F-theory,
nor the heterotic string is expected to provide a good
description. For example, consider the $E_8\times E_8$ heterotic string with
$(12-n, 12+n)$ instantons embedded in the gauge group. This model has a dual
description in terms of F theory compactified on a Calabi-Yau three-fold with
base $F_n$ \rFTHEORY. The base of the three-fold is a $P^1$ fibered over a
$P^1$, with a twist determined by the integer $n$. The heterotic coupling is
given by the ratio
of the volumes of the two $P^1$ factors. If we choose to keep
this ratio of order one, then the perturbative heterotic string is no longer a
good description of the physics. F theory provides a good description when the
size of both $P^1$ factors is large. When both spheres have size of order
unity, we expect the F theory description
to receive significant corrections. However, when the F theory
compactification
can be related to an orientifold theory, the
orientifold should
continue to provide a good description of the theory
in this region as long as the coupling
constant in the orientifold theory remains small. If we can
apply T-duality to the orientifold description to convert it into a type I
compactification, we
should find a description in terms of
weakly coupled type I
string theory for this region of the moduli space.
Compactifications to four dimensions have a
correspondingly richer class of degenerations, where analogous questions arise.

The cases where T-duality has been used to relate an orientifold to a type I
theory have primarily been orientifolds of type IIB compactified on a torus.
The aim of this paper is to relate certain compactifications of type I on $K3$
to
 orientifolds of type IIB on $K3$, where the $K3$ is not necessarily a toroidal
orbifold. The gauge bundle of the type I theory falls into one of three
topologically distinct classes, as described in \rWITSIX. In the following
section, we relate these type I vacua to IIB orientifolds which may have a
non-trivial $B_{\mu\nu}$ background flux, with the quotient group generated
by $\Omega$. To connect the type I vacua to F
theory, we will want to get rid of the background flux, and also have the
quotient group generated by $(-1)^{F_L}\cdot\Omega$, accompanied by a
geometric transformation on $K3$. In section three, we
describe how to use a mirror transformation to achieve both these goals.
In turn, these models are in the moduli space of F
theory compactifications on three-folds with base $F_n$, where $n=0,1,4$. The
case with $n=2$ is known to be connected to the
$n=0$ model \refs{\rSW,\rFTHEORY}. In this way,
we will connect the type I models with these
particular F theory compactifications,  and their corresponding
dual $E_8\times E_8$ heterotic
string compactifications. Aspinwall has recently obtained similar results using
quite different techniques \rASPINEW.

\newsec{Orientifolds and Type I Vacua}

To build a compactification of the Spin(32)/$Z_2$ heterotic or
type I string theory
on $K3$, we need to specify a Spin(32)/$Z_2$ gauge bundle. The instanton number
of the bundle together with the number of five-branes filling space-time must
be 24 in order
to satisfy anomaly cancellation. It has been shown in \rWITSIX, and
subsequently in \rASPINEW,
that the Spin(32)/$Z_2$ gauge bundle on $K3$ can belong
to one of three topological classes,
characterized by an element $\wt w_2$ of $H^2(K3,Z)$, known as the
``generalized second Stiefel-Whitney class.'' Actually, since
$\wt w_2$ is defined modulo a shift by twice a lattice vector of
$H^2(K3,Z)$, it is really an element of $H^2(K3, Z_2)$. We will usually view
$\wt w_2$ as an integer class, which is well-defined mod 2. As in
\rASPINEW, we shall find it convenient to use Poincare duality
to represent $\wt w_2$ as an element of $H_2(K3,Z)$. The physical
significance of $\wt w_2$ can be described as follows. Let $C$ denote a
two sphere in $K3$, and $C_N$ and $C_S$ denote its northern and
southern hemispheres.  Let $g_N$ and $g_S$ denote the holonomies along
the boundaries of $C_N$ and $C_S$ in the {\it vector representation} of
$SO(32)$. Since the boundaries of $C_N$ and $C_S$ denote the same curves,
$g_Ng_S^{-1}$ must be unity in Spin(32)/$Z_2$, but not necessarily so
in $SO(32)$. $\wt w_2$ associated with the gauge bundle is defined
such that in the vector representation of $SO(32)$,
\eqn\1{g_N g_S^{-1} = \exp (i\pi {\wt w_2} \cdot C),}
where $ \cdot $ denotes intersection number. Just as the usual second
Stiefel-Whitney class describes an obstruction to a bundle admitting spinors,
${\wt w_2}$ describes an obstruction to a bundle admitting vectors. It is worth
recalling that $H_2(K3, Z)$ is an even, self-dual lattice. Up to
diffeomorphism, there are then three independent topological
classes of gauge bundles, which correspond to the cases \refs{\rWITSIX,
\rASPINEW},
\vskip 0.15in
\item{1.}{${\wt w_2} =0,$}
\vskip 0.02in
\item{2.}{${\wt w_2}\cdot{\wt w_2} =0$ mod 4, and}
\vskip 0.02in
\item{3.}{${\wt w_2}\cdot{\wt w_2} =2$ mod 4.}
\vskip 0.15in\noindent
The first case corresponds to a conventional $SO(32)$ bundle, while the second
is pertinent to the models described in \rPOL, when the fixed-points are blown
up. We will meet all three cases in the following discussion.

Using a well known fact, we can regard the type I theory on $K3$ as type IIB
theory on $K3$ modded out by the world-sheet parity transformation,
$\Omega$. In order to be able to mod out the type IIB theory by $\Omega$,
we must ensure that the original field configuration is invariant under
$\Omega$. Since the rank two anti-symmetric tensor $B_{\mu\nu}$
originating in the Neveu-Schwarz Neveu-Schwarz sector is odd under $\Omega$,
one might naively think that the components of $B_{\mu\nu}$
along $K3$ need to be set to zero in order to get an $\Omega$ invariant
configuration. However, this is not strictly
necessary. To see this, note that the components of $B$ along $K3$
belong to $H^2(K3,R)$ modulo elements
of $H^2(K3,Z)$ because of the periodicity of the flux of $B_{\mu\nu}$.
Using Poincare duality we can take these to be elements
of  $H_2(K3,R)$ modulo elements of
$H_2(K3,Z)$. In this notation, if $B$ is $\Lambda/2$ for some element
$\Lambda$ of $H_2(K3,Z)$, then it is invariant under $\Omega$, since
$\Lambda/2$ and $-\Lambda/2$ are identified under the action of
$H_2(K3,Z)$.  Thus, we can mod out such a configuration of type IIB theory
on $K3$ by $\Omega$ to get an unconventional type I theory on $K3$. The
relation between quantized anti-symmetric tensor field flux and unbroken gauge
symmetry has previously been noted in \rBPS.
Since in this theory, such a $B$-flux cannot be continuously deformed to
zero, we shall refer to this as a
discrete $B$-flux. We shall now
show that the presence of such a flux
actually leads us to a type I theory with $\wt w_2=\Lambda$.

Let $C$  be any two sphere inside $K3$, and let us consider a world-sheet
instanton in type I theory, which represents an elementary string world-sheet
with
spherical topology wrapped once around $C$. In the
presence of a $B$-flux, $\Lambda/2$, through the two-cycles of $K3$, the
phase associated with such an instanton  will be given by,
\eqn\2{
\exp(i\pi\Lambda\cdot C).
}
This instanton contribution represents an amplitude for a closed string to be
produced from the vacuum, then propagate for a finite interval of time during
which it wraps around $C$, and finally disappear into the vacuum.  Now,
since in type I string theory, a closed string carries the same quantum number
as a pair of open strings with their ends glued to each other, we should be
able to calculate this phase by considering the propagation of a pair of
open strings, spanning the northern and the southern hemispheres
$C_N$ and $C_S$ respectively. This corresponds to replacing the sphere by
two disks, spanning $C_N$ and $C_S$. The phases associated with these disk
amplitudes are given by $g_N$ and $g_S^{-1}$ respectively.\foot{We
get $g_S^{-1}$ instead of $g_S$ since the ends of the open string
spanning $C_S$ carry charge opposite to that spanning $C_N$.} Thus, the
total phase is given by $g_N g_S^{-1}$. This, in turn, is given
by \1. Comparing \1\ and \2, we now find that
\eqn\3{
\wt w_2 = \Lambda,
}
up to a shift by twice a lattice vector in $H_2(K3,Z)$.

\newsec{The Mirror Transformation}

Now that we have identified a type IIB orientifold that corresponds to a type I
compactification with non-trivial ${\wt w_2}$, we want to dualize the model to
obtain an orientifold without any discrete $B$-flux,
and with the
orientifold group generated by $(-1)^{F_L}\cdot\Omega$ together with
an involution on the new $K3$. For this we need
to review a few facts about the conformal field theory describing propagation
of type II strings
on $K3$. Let us recall that
$H_2(K3,Z)$ has the structure of an even self-dual lattice $\Gamma^{(3,19)}$
with
signature (3,19), which can be decomposed in the following way:
\eqn\AAA{\Gamma^{(3,19)}=3\,\Gamma^{(1,1)} \oplus 2\,\Gamma^{(E_8)}.
}
In the conformal field theory which describes type II strings on $K3$, it is
natural to
introduce a lattice $\Gamma^{(4,20)}$ of signature (4,20) by adding an extra
copy of $\Gamma^{(1,1)}$ to $\Gamma^{(3,19)}$:
\eqn\EE{\Gamma^{(4,20)}=\Gamma^{(1,1)}\oplus \Gamma^{(3,19)},
}
where this extra $\Gamma^{(1,1)}$ can be associated with the 0- and the
4-cycles of $K3$ \rASPMOR. Locally,
the moduli space of this conformal field theory can be 
identified with the Grassmannian of 4-planes of signature (4,0) in
$R^{(4,20)}$, and is parametrized by an
$O(4,20)$ matrix $N$ modulo multiplication from the right by an $O(4)\times
O(20)$ matrix \rASPINK.  
For a given $N$, we obtain a 4-plane of signature 
(4,0) in this twenty-four dimensional space by using the projection
operator:
\eqn\FF{
{1\over 2}(LNN^T + I),
}
where $L$ is the metric of signature
(4,20)  and $I$ is the $24\times 24$ identity matrix.
To see that this is a 
projection operator, note that by definition, $ N^T L N = L$.
After rotating the lattice using the $O(4,20)$ matrix $N^{T}$, we can turn
this projection operator into the operator $(I+L)$, which makes it clear that
the operator \FF\ defines a (4,0) subspace \rNSW.
We shall find it convenient to use this
description.

Using the freedom of multiplying $N$ by an $O(4)\times O(20)$
matrix from the right, we can choose $N$ to be of the form:
\eqn\CCC{
N=N_BN_SN_K,
}
where $N_K$ is an element of $O(3,19)$ that rotates the basis vectors
of $\Gamma^{(3,19)}$, $N_S$
is an element of $O(1,1)$ that rotates the basis vectors of
$\Gamma^{(1,1)}$, and $N_B$ can be represented as an
upper triangular matrix\foot{In order to represent $N_B$ 
as an upper triangular matrix we need to choose the two basis vectors of
$\Gamma^{(1,1)}$ as the first and the last row/column, with the basis
vectors of $\Gamma^{(3,19)}$ labelling the rest of the rows and columns.
Later in eq. (3.8),
we shall construct this matrix explicitly for a given $B$-flux.}
that mixes the basis vectors of $\Gamma^{(1,1)}$ with
the basis vectors of $\Gamma^{(3,19)}$.
Then the parameter labelling $N_S$
can be identified as the size of $K3$, those labelling $N_K$
correspond to the rest of the geometric moduli of $K3$, while the parameters
appearing in $N_B$ characterize the $B$-flux.

Let us now choose a $\Gamma^{(1,1)}$ component of the lattice
$\Gamma^{(3,19)}$ that is orthogonal to $\Lambda$.
In order not to confuse this with the part of the lattice associated with
the 0- and 4- cycles, we shall denote this by $\Gamma^{(1,1)\prime}$.
Orthogonality of $\Lambda$ to $\Gamma^{(1,1)\prime}$ guarantees that
$N_B$ does not act on $\Gamma^{(1,1)\prime}$. Furthermore, by
continuously adjusting the geometric moduli
of $K3$, we can ensure that $N_K$
does not act on $\Gamma^{(1,1)\prime}$. In that case
$N$ leaves $\Gamma^{(1,1)\prime}$ invariant. Now, let us make a mirror
transformation that converts the pair of two cycles associated with
$\Gamma^{(1,1)\prime}$ into the zero and four cycles of the mirror $K3$,
which we shall denote by $K3'$.
In this case invariance of $\Gamma^{(1,1)\prime}$
under $N$ can be interpreted as the absence of any background $B$-flux
in this mirror transformed model. If we had started from a generic
point in the moduli space of the original $K3$ then after the mirror
transformation, we would have got a $K3'$ with $B$-flux, but this flux will not
be discrete in the sense described earlier,
since it can be continuously
deformed to zero, while maintaining invariance under the mirror image of
$\Omega$.
Hence, we have arrived at a model without any discrete $B$-flux.

This mirror transformation
converts the four cycle and the zero cycle of the original $K3$ into
a two cycle and its dual two cycle of
the mirror $K3'$.
Studying the action of various discrete symmetries on the massless fields
in the theory, one can easily verify that the mirror transformation
conjugates $\Omega$ to
\eqn\4{
(-1)^{F_L}\cdot\Omega\cdot\sigma,
}
where $\sigma$ is a Nikulin
involution of $K3'$ that leaves this particular pair of
two cycles invariant, but changes the sign of all other
two cycles \rNIKULIN.\foot{To see this, note that 
in the original six dimensional theory, the only rank two
anti-symmetric tensor fields
invariant under $\Omega$ come from the Ramond-Ramond antisymmetric tensor
field $B'_{\mu\nu}$ and its magnetic dual.
Under the mirror
transformation these get mapped to the components of the rank two
anti-symmetric tensor field $D_{\mu\nu\rho\sigma}$ along this
particular pair of two cycles in $K3'$. Since $D_{\mu\nu\rho\sigma}$ is
even under $(-1)^{F_L}\cdot\Omega$, $\sigma$ must leave this particular
pair of two cycles invariant and change the sign of all other two cycles.}
The question we shall be
addressing is the following:
if we start from a  particular $B$ flux background
$\Lambda/2$ in the
original type IIB theory, then which particular Nikulin involution
do we get after the mirror transformation?

Let us recall that a Nikulin involution is characterized by three
integers $(r,a,\delta)$ defined as follows.  If $S_+$ denotes
the sublattice of $H_2(K3',Z)$ that is invariant under $\sigma$,  and
$S_+^*$ denotes the dual lattice of $S_+$, then

\vskip 0.15in
\item{1.}{$r$ denotes the rank of $S_+$,}
\vskip 0.02in
\item{2.}{$a$ denotes that $S^*_+/S_+$ has the structure of $(Z/2Z)^a$ and,}
\vskip 0.02in
\item{3.}{$\delta=0$ if $x^2$ is integer for all $x\in S^*_+$, otherwise
$\delta=1$.}
\vskip 0.15in\noindent

For our purpose, it will be useful to translate this into a statement
about the action of $(-1)^{F_L}\cdot\Omega\cdot\sigma$ on the lattice of
allowed charges under the rank two tensor gauge fields arising in
the Ramond-Ramond (RR) sector of the theory. There are twenty-four such
(anti-)self-dual tensor fields coming from the RR sector, of which twenty-two
come from the components of the rank four RR gauge field
$D_{\mu\nu\rho\sigma}$ along the two-cycles of $K3'$. The charges associated
with these fields clearly belong to the lattice $H_2(K3',Z)$. The other
two come from the RR 
two-form gauge field $B'_{\mu\nu}$ and its magnetic dual.
The allowed charges associated
with these fields belong to the self-dual Lorentzian lattice
$\Gamma^{(1,1)\prime}$. Together they span a lattice $\Gamma^{(4,20)\prime}$,
which of course is isomorphic to the the lattice of charges $\Gamma^{(4,20)}$
of the original type IIB theory on $K3$.

Now the RR four-form field is invariant under $(-1)^{F_L}\cdot\Omega$. On the
other hand,
the RR two-form field is odd under this transformation, since it is odd under
$(-1)^{F_L}$ but invariant under $\Omega$. Thus acting on
the lattice of tensor field charges, $(-1)^{F_L}\cdot\Omega\cdot\sigma$ acts as
$\sigma$ on the $H_2(K3',Z)$ part of the lattice, and reverses
the sign of the $\Gamma^{(1,1)\prime}$ part of the lattice. Thus $S_+$, which
was earlier defined as the sublattice of $H_2(K3',Z)$ invariant under $\sigma$,
can also be regarded as the sublattice of $\Gamma^{(4,20)\prime}$ invariant
under $(-1)^{F_L}\cdot\Omega\cdot\sigma$.

So far we have not learned anything new, but now we shall use the
fact that a mirror transformation acts as an automorphism of the charge
lattice \rASPMOR. Thus $S_+$ can also
be regarded as the sublattice of $\Gamma^{(4,20)}$ invariant under $\Omega$
in the original type IIB theory, since $\Omega$, by construction,
is the image of $(-1)^{F_L}\cdot\Omega\cdot\sigma$ under the mirror map.
This allows us to forget all reference to the mirror $K3'$ and work
with the original type IIB theory on $K3$.
First let us consider the case where there is
no $B$ flux in the original theory.
Since $\Omega$ changes the sign of the rank four gauge-field, but
leaves the rank two RR tensor $B'_{\mu\nu}$ invariant, its action on the
charge lattice will be to change the sign of the $H_2(K3,Z)$ part of the
lattice, leaving the $\Gamma^{(1,1)}$ part associated with the zero and
four cycles invariant. Thus we can identify
$S_+$ with $\Gamma^{(1,1)}$. Since $S_+$ is an even self-dual lattice of
signature (1,1), we find, from the definitions of $(r,a,\delta)$, that in
this case,
\eqn\5{
(r,a,\delta) = (2,0,0).
}
Let us now consider the case where there is
a background $B$-flux equal
to $\Lambda/2$, for  some element $\Lambda$ of $H_2(K3,Z)$. Let
$e_1$ and $e_2$ denote the basis vectors of $\Gamma^{(1,1)}$
associated with the zero and the four cycles ({\it i.e.} $B'_{\mu\nu}$
charges) with the inner product matrix:
\eqn\6{
e_1^2=e_2^2=0, \quad e_1\cdot e_2=1.
}
Also let $\Lambda$, together with a set of vectors $f_i$, be the generators
of $H_2(K3,Z)$. Note that since $\Lambda$ is defined modulo a shift by twice a
lattice
vector, we can always choose $\Lambda$ to be a primitive vector
of the lattice. The $O(4,20)$ matrix $N_B$ 
representing the effect of this $B$-flux induces the following rotation of
the basis vectors,
\eqn\7{\eqalign{
e_1' &= e_1, \cr
e_2' &= e_2 + {1\over 2}\Lambda -{1\over 8} \Lambda^2 e_1,\cr
\Lambda' &= \Lambda - {1\over 2} \Lambda^2 e_1, \cr
f_i' &= f_i -{1\over 2} (f_i\cdot\Lambda) e_1.
}}

This rotation preserves the inner products between different basis vectors,
reflecting the fact that it is an $O(4,20)$ rotation.
The charge vectors lie on the new lattice generated by 
the primed vectors; however $B'_{\mu\nu}$, its magnetic dual, and the
components of $D_{\mu\nu\rho\sigma}$ along the two cycles of $K3$
continue to couple to the components of the charge vector along
the unprimed vectors $e_1$, $e_2$ and $\{\Lambda,f_i\}$ respectively.
The vectors $e_1'$ and $e_2'$ generate a
$\Gamma^{(1,1)}$ component of the new lattice. However, the action
of $\Omega$ no longer leaves this $\Gamma^{(1,1)}$ invariant, since it
leaves $e_1$ and $e_2$ ({\it not} $e_1'$ and $e_2'$) invariant,
changing the signs of $\Lambda$ and the $f_i$'s. In other words, if we
consider an element of the new $\Gamma^{(4,20)}$ lattice of the form:
\eqn\8{
n_1 e_1'+n_2 e_2' + n\Lambda'+ m_i f_i',
}
with $n_i$, $m_i$, $n$ integers, then invariance under $\Omega$ requires that
this vector must lie in the $e_1$-$e_2$ plane. Using \7, this gives
the following constraints:
\eqn\9{
m_i=0, \qquad n +{1\over 2} n_2=0.
}
Using equations \7\ and \9, we can rewrite \8\ as,
\eqn\A{
(n_1 -{1\over 4} n \Lambda^2) e_1 - 2n e_2.
}
Therefore, $S_+$ is the lattice of vectors of the form given in \A,
for integer $n$ and $n_1$. Clearly this has rank two, so here $r=2$. With
the help of the inner product \6\ and the fact that $\Lambda^2$ is
even, we see that a general element of the dual lattice $S_+^*$ is of the form:
\eqn\B{
{1\over 2} (k_1 - {1\over 4}k \Lambda^2)e_1 - ke_2,
}
for integer $k_1$ and $k$. Comparing \A\ and \B, we
see that $S_+^*/S_+$ is isomorphic to $(Z/2Z)^2$. This gives $a=2$. Finally,
the inner product of the element \B\ with itself is given by
\eqn\C{
-k (k_1 -{1\over 4} k \Lambda^2).
}
Thus if $\Lambda^2=0$ mod 4, this inner product is always integer and we
have $\delta=0$. On the other hand, if $\Lambda^2=2$ mod 4, then this
inner product is half integer for odd $k$, and we have $\delta=1$.

Using the identification of $\Lambda$ with $\wt w_2$, we now have the
following identification between type I compactification on $K3$
and type IIB on  $K3/(-1)^{F_L}\cdot\Omega\cdot\sigma$, with $\sigma$
denoting Nikulin involution $(r,a,\delta)$:

\vskip 0.15in
\item{1.}{$\wt w_2=0$: \quad $(r,a,\delta)$=(2,0,0).}
\vskip 0.02in
\item{2.}{$\wt w_2\cdot\wt w_2=0$ mod 4: \quad $(r,a,\delta)$=(2,2,0).}
\vskip 0.02in
\item{3.}{$\wt w_2\cdot\wt w_2=2$ mod 4: \quad $(r,a,\delta)$=(2,2,1).}
\vskip 0.15in\noindent

The orientifolds appearing in the above relations are known to be
dual to F-theory on elliptically fibered Calabi-Yau 3-folds on
base $F_4$, $F_0$ and $F_1$, respectively \refs{\rFTHEORY,\rFGEN}. Had we
arrived at any other Nikulin actions and hence different orientifolds, we would
certainly have been in trouble. The type I models with which we started had a
single tensor multiplet, but F theory compactified
on other Voisin-Borcea spaces
has more (or in some cases less) than one tensor multiplet.
The particular F theory models that we have found are, in turn,
dual to the $E_8\times E_8$ heterotic string compactified on $K3$,
with the 24 instantons embedded in the two $E_8$ factors of the gauge group
according to the distribution (8,16),
(12,12) and (11,13), respectively \rFTHEORY. Finally,
using T-duality between the two heterotic string theories, these models
can be identified with the Spin(32)/$Z_2$ heterotic string theory on $K3$,
with $\wt w_2=0$, $\wt w_2\cdot\wt w_2=0$ mod 4 and $\wt w_2
\cdot\wt w_2=2$ mod 4, respectively \refs{\rWITSIX, \rASPINEW}. Therefore, we
see that, using the T-duality relations we have discovered,
we have travelled a full
circle in duality transformations via the route type I - orientifold -
F-theory - $E_8\times E_8$ heterotic - Spin(32)/$Z_2$
heterotic - type I  
compactifications.

\bigbreak\bigskip\bigskip\centerline{{\bf Acknowledgements}}\nobreak

It is a pleasure to thank R. Gopakumar and E. Witten for helpful
discussions, and
J. Schwarz for a careful reading of the manuscript.
The work of A.S. is supported by a grant from NM Rothschild
and sons Ltd, while that of S.S. is
supported by NSF grant DMS--9627351.

\vfill\eject

\listrefs
\bye